\documentclass{aastex61}

\received{}
\revised{}
\accepted{}
\submitjournal{ApJ}

\shorttitle{Qatar-3b, Qatar-4b, Qatar-5b exoplanets}
\shortauthors{Aslubai et al.}

\usepackage{amsmath,texlogos}
\usepackage{natbib}
\usepackage{subfigure}
\usepackage{graphicx}
\usepackage{txfonts}
\usepackage[T1]{fontenc}
\usepackage{aecompl}
\usepackage{times}

\newcommand{\kms}{\ensuremath{\rm km\,s^{-1}}}
\newcommand{\ms}{\ensuremath{\rm m\,s^{-1}}}
\newcommand{\logg}{\ensuremath{\log{g}}}
\newcommand{\vsini}{\ensuremath{v\,\sin{i}}}

\newcommand{\bjdtdb}{\ensuremath{\rm {BJD_{TDB}}}}
\newcommand{\feh}{\ensuremath{\left[{\rm Fe}/{\rm H}\right]}}
\newcommand{\mh}{\ensuremath{\left[{\rm M}/{\rm H}\right]}}
\newcommand{\teff}{\ensuremath{T_{\rm eff}}}

\newcommand{\msun}{\ensuremath{\,M_\odot}}
\newcommand{\rsun}{\ensuremath{\,R_\odot}}
\newcommand{\lsun}{\ensuremath{\,L_\odot}}
\newcommand{\mj}{\ensuremath{\,M_{\rm J}}}
\newcommand{\rj}{\ensuremath{\,R_{\rm J}}}
\newcommand{\fave}{\langle F \rangle}
\newcommand{\fluxcgs}{10$^9$ erg s$^{-1}$ cm$^{-2}$}

\begin{document}

\title{Qatar Exoplanet Survey : Qatar-3b, Qatar-4b and Qatar-5b.}

\correspondingauthor{Khalid Alsubai}
\email{kalsubai@qf.org.qa}

\author[0000-0002-0786-7307]{Khalid Alsubai}
\affil{Qatar Environment and Energy Research Institute (QEERI), HBKU, Qatar Foundation, PO Box 5825, Doha, Qatar}

\author{Dimitris Mislis}
\affiliation{Qatar Environment and Energy Research Institute (QEERI), HBKU, Qatar Foundation, PO Box 5825, Doha, Qatar}

\author{Zlatan I. Tsvetanov}
\affiliation{Qatar Environment and Energy Research Institute (QEERI), HBKU, Qatar Foundation, PO Box 5825, Doha, Qatar}

\author{David W. Latham}
\affiliation{Harvard-Smitsonian Center for Astrophysics, 60 Garden Street,  Cambridge, MA 02138, USA}

\author{Allyson Bieryla}
\affiliation{Harvard-Smitsonian Center for Astrophysics, 60 Garden Street,  Cambridge, MA 02138, USA}

\author{Lars A. Buchhave}
\affiliation{Centre for Star and Planet Formation, Natural History Museum of Denmark \& Niels Bohr Institute, University of Copenhagen,{\O}ster Voldgade 5-7, DK-1350 Copenhagen K, Denmark}

\author{Gilbert A. Esquerdo}
\affiliation{Harvard-Smitsonian Center for Astrophysics, 60 Garden Street,  Cambridge, MA 02138, USA}

\author{D. M. Bramich}
\affiliation{Qatar Environment and Energy Research Institute (QEERI), HBKU, Qatar Foundation, PO Box 5825, Doha, Qatar}

\author{Stylianos Pyrzas}
\affiliation{Qatar Environment and Energy Research Institute (QEERI), HBKU, Qatar Foundation, PO Box 5825, Doha, Qatar}

\author{Nicolas P. E. Vilchez}
\affiliation{Qatar Environment and Energy Research Institute (QEERI), HBKU, Qatar Foundation, PO Box 5825, Doha, Qatar}

\author{Luigi Mancini}
\affiliation{Max Planck Institute for Astronomy, K\"{o}nigstuhl 17, D-69117 Heidelberg, Germany}
\affiliation{INAF-Osservatorio Astrofisico di Torino, via Osservatorio 20, 10025, Pino Torinese, Italy}

\author{John Southworth}
\affiliation{Astrophysics Group, Keele University, Staffordshire ST5 5BG, UK}

\author{Daniel F. Evans}
\affiliation{Astrophysics Group, Keele University, Staffordshire ST5 5BG, UK}

\author{Thomas Henning}
\affiliation{Max Planck Institute for Astronomy, K\"{o}nigstuhl 17, D-69117 Heidelberg, Germany}

\author{Simona Ciceri}
\affiliation{Max Planck Institute for Astronomy, K\"{o}nigstuhl 17, D-69117 Heidelberg, Germany}
\affiliation{Department of Astronomy, Stockholm University, AlbaNova University Centre, SE-106 91 Stockholm, Sweden}



\begin{abstract}

We report the discovery of Qatar-3b, Qatar-4b, and Qatar-5b, three new transiting planets 
identified by the Qatar Exoplanet Survey (QES). The three planets belong to the hot Jupiter 
family, with orbital periods of $P_{Q3b}$\,=\,2.50792 days, $P_{Q4b}$\,=\,1.80536 days, 
and $P_{Q5b}$\,=\,2.87923 days. Follow-up spectroscopic observations reveal the masses 
of the planets to be $M_{Q3b}$\,=\,4.31$\pm0.47$\,$M_{\rm J}$, $M_{Q4b}$\,=\,5.36$\pm0.20$\,$M_{\rm J}$, 
and $M_{Q5b}$\,=\,4.32$\pm0.18$\,$M_{\rm J}$, while model fits to the transit light curves yield radii 
of $R_{Q3b}$\,=\,1.096$\pm0.14$\,$R_{\rm J}$, $R_{Q4b}$\,=\,1.135$\pm0.11$\, $R_{\rm J}$, and 
$R_{Q5b}$\,=\,1.107$\pm0.064$\, $R_{\rm J}$. The host stars are low-mass main 
sequence stars with masses and radii $M_{Q3}$\,=\,1.145$\pm0.064$\,$M_{\odot}$, 
$M_{Q4}$\,=\,0.896$\pm0.048$\,$M_{\odot}$, $M_{Q5}$\,=\,1.128$\pm0.056$\,$M_{\odot}$ and 
$R_{Q3}$\,=\,1.272$\pm0.14$\,$R_{\odot}$, $R_{Q4}$\,=\,0.849$\pm0.063$\,$R_{\odot}$ and 
$R_{Q5}$\,=\,1.076$\pm0.051$\,$R_{\odot}$ for Qatar-3, 4 and 5 respectively. The V magnitudes 
of the three host stars are $V_{Q3}$=12.88, $V_{Q4}$=13.60, and $V_{Q5}$=12.82. All three new 
planets can be classified as heavy hot Jupiters (M > 4 $M_{J}$).


\end{abstract}




\section{Introduction}

Ground-based surveys for transiting exoplanets continue to be a productive source for 
finding short period giant planets orbiting relatively bright stars. Many of these discoveries 
have become primary targets for subsequent studies of exoplanetary atmospheres and 
other important planetary characteristics with the use of some of the most advanced 
ground- and space-based telescopes. In addition, these discoveries contribute 
significantly to a more complete census of hot Jupiters and other close orbiting large 
planets -- the type of planets not present in our solar system -- and may provide a key 
to understanding their origin and more generally the planetary system architecture.

This paper is based on observations collected with the first generation of the Qatar 
Exoplanet Survey (QES, \citealt{alsubai}). QES uses two overlapping wide field 135mm 
(f/2.0) and 200mm (f/2.0) telephoto lenses, along with four 400mm (f/2.8) telephoto 
lenses, mosaiced to image an $11^{\rm o}\,\times\,11^{\rm o}$ field on the sky simultaneously 
at three different pixel scales. The three different pixel scales are 12, 9 and 4 arcsec
respectively for the three different type of lenses. With its larger aperture lenses, its 
higher angular resolution (a result of the longer focal length of the lenses), and the detrending 
algorithms, QES is able to reach 1\% photometric accuracy up to 13.5-14.0 mags.  

In this paper we present the discovery of three new hot Jupiters from QES, namely Qatar-3b, 
Qatar-4b, and Qatar-5b. The paper is organized as follows: in section \ref{sec:obser} we 
present the survey photometry and describe the follow-up spectroscopy and  photometry 
used to confirm the planetary nature of the transits. In section \ref{sec:anares} we present 
the global system solutions using simultaneous fits to the available RV and follow-up 
photometric light curves with the stellar parameters determined from the combined spectra, 
while in Section 4 we summarise our results. 

\section{Observations} \label{sec:obser} 

\subsection{Discovery photometry} \label{subsec:discphot}

Observations for the discovery photometry were collected at the QES station 
in New Mexico, USA. QES utilizes FLI ProLine PL6801 cameras, with 
KAF-1680E 4k$\times$4k detectors. Exposure times were 60s, for each of the 
four CCDs attached to the 400mm lenses; 45s, for the CCD equipped with the 
200mm lens; and 30s, for the CCD equipped with the 135mm lens. 

The survey data were reduced with the QES pipeline, which performs 
bias-correction, dark-current subtraction and flat-fielding in the standard 
fashion, while photometric measurements are extracted using the image 
subtraction algorithm by \cite{dbdia}; a more detailed description of the 
pipeline is given in \cite{alsubai}.

\begin{figure}
\centering
\includegraphics[width=8.5cm]{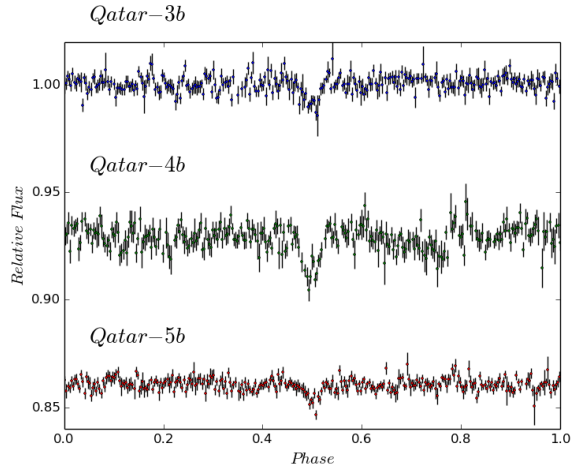} 
\caption{The discovery light curves phase folded with the BLS estimated 
periods, as they appear in the QES archive, for Qatar-3b (top, in blue), 
Qatar-4b (middle, in green), and Qatar-5b (bottom, in red) [color available 
in the on-line version only]. For clarity, all light curves have been binned using 
a mean filter by a factor of 35, while those of Qatar-4b and Qatar-5b
have been shifted downwards.}
\label{fig1}
\end{figure}

The output light curves were ingested into the QES archive and subsequently 
subjected to a combination of the Trend Filtering Algorithm (TFA, \citealt{kovacs1}) 
and the SysRem algorithm \citep{tamuz}, to model and remove systematic patterns 
of correlated noise. Transit-like events for all three stars were identified 
using the Box Least Square algorithm (BLS) of \cite{kovacs2}, during a candidates'
search on the archive light curves following the procedure described in \cite{collier}. We note 
that the initial candidate selection is an automatic procedure, but the final candidate vetting 
is done by eye. The BLS algorithm provided tentative ephemerides which were 
used to phase-fold the discovery light curves shown in Figure\,\ref{fig1}. 

The discovery light curve of Qatar-3b contains data points from 11,228 frames, 
spanning a period from October 2012 to January 2015, that of Qatar-4b contains 
data points from 8,950 frames, with a time-span from September 2012 to 
November 2014, and that of Qatar-5b contains 18,957 data points, with a 
time-span from September 2012 to December 2014. 


\subsection{The host stars} \label{subsec:hosts}

Qatar-3b's host is a V\,=\,12.88 mag (B\,=\,13.13 mag) star (UCAC3 ID: 253-304972, 
henceforth designated Qatar-3) of spectral type very close to G0V. The host of Qatar-4b 
is a V\,=\,13.60 mag (B\,=\,14.69 mag), early-K type star (UCAC3 ID: 269-003518, 
henceforth designated Qatar-4), and, similarly, the host of Qatar-5b is a V\,=\,12.82 
mag (B\,=\,13.00 mag) star (UCAC3 ID: 265-004681, henceforth designated 
Qatar-5) of spectral type close to G2V. The basic observational characteristics of the 
three host stars, together with the results from the spectroscopic analysis, are listed in 
Table \ref{tab:t1}. We further discuss stellar parameters determined from our follow-up 
spectra in section \ref{subsec:SPC}. The host star spectral types are estimated from a 
multi-color fit (J, H, V and K band) to the UCAC3 values, using a standard Random-Forest 
classification algorithm, trained with $\sim$200 standards with spectral types ranging 
from early A to late M.

\begin{table}
\centering
\caption{\label{tab:t1} Basic observational and spectroscopic parameters of the host stars}
\begin{tabular}{lccc}
\hline
Parameters & Qatar-3         & Qatar-4        & Qatar-5 \\ 
\hline
ID (UCAC3) & 253-304972 & 269-003518 & 265-004681 \\
$\alpha_{\mathrm{2000}}$  & $23^{\mathrm{h}}56^{\mathrm{m}}36.48^{\mathrm{s}}$ & $00^{\mathrm{h}}19^{\mathrm{m}}26.22^{\mathrm{s}}$ & $00^{\mathrm{h}}28^{\mathrm{m}}12.94^{\mathrm{s}}$ \\    
$\delta_{\mathrm{2000}}$ & $+36^{\mathrm{o}}12^{\prime}46.6^{\prime\prime}$ & $+44^{\mathrm{o}}01^{\prime}39.4^{\prime\prime}$ & $+42^{\mathrm{o}}03^{\prime}40.9^{\prime\prime}$ \\
V [mag] & 12.88 & 13.60 & 12.82 \\
B [mag] & 13.13 & 14.69 & 13.00\\
J [mag] & 11.60 & 13.61 & 11.35\\
Spectral Types & G0V & K1V & G2V \\
                &           &           &   \\
\teff\, [K]        & 5991$\pm$64      & 5198$\pm$42     & 5746$\pm$50  \\
\logg\, [cgs]   &  4.28$\pm$0.05   & 4.56$\pm$0.06   & 4.43$\pm$0.10 \\
\mh               & -0.02$\pm$0.07   & 0.14$\pm$0.09   & 0.38$\pm$0.08 \\
\vsini\, [\kms] & 10.4$\pm$0.5     &   7.1$\pm$0.3     & 4.5$\pm$0.5 \\

\hline
\end{tabular}
\end{table}

\subsection{Follow-up spectroscopy} \label{subsec:spec}

Follow-up spectroscopic observations of all three candidates were obtained 
with the Tillinghast Reflector Echelle Spectrograph (TRES) on the 1.5\,m 
Tillinghast Reflector at the Fred L.\,Whipple Observatory on Mount Hopkins, 
Arizona. Similarly to our campaigns for all QES candidates we used TRES 
with the medium fiber, which yields a resolving power of $R \sim$ 44,000, 
corresponding to a velocity resolution element of 6.8 \kms\ FWHM. The 
spectra were extracted using version 2.55 of the code described in 
\cite{buchhave2010}. The wavelength calibration for each spectrum was 
established using exposures of a thorium-argon hollow-cathode lamp 
illuminating the science fiber, obtained immediately before and after each 
observation of the star. 

For Qatar-3 a total of 34 spectra were obtained between 2015-07-30 (UT) 
and 2016-10-24 with a typical exposure time of 30 min and an average 
signal-to-noise ratio per resolution element (SNRe) of 29 at the peak of the 
continuum in the echelle order centered on the Mg b triplet near 519 nm. 
For Qatar-4 we obtained 10 usable spectra between 2015-09-23 and 
2017-01-19 with mostly 48-min exposures and <SNRe>=22, and for 
Qatar-5 a total of 25 usable spectra between 2015-09-27 and 2015-12-08 
with mostly 25-min exposures and <SNRe>=29.

\begin{figure}
\centering
\includegraphics[width=8.5cm]{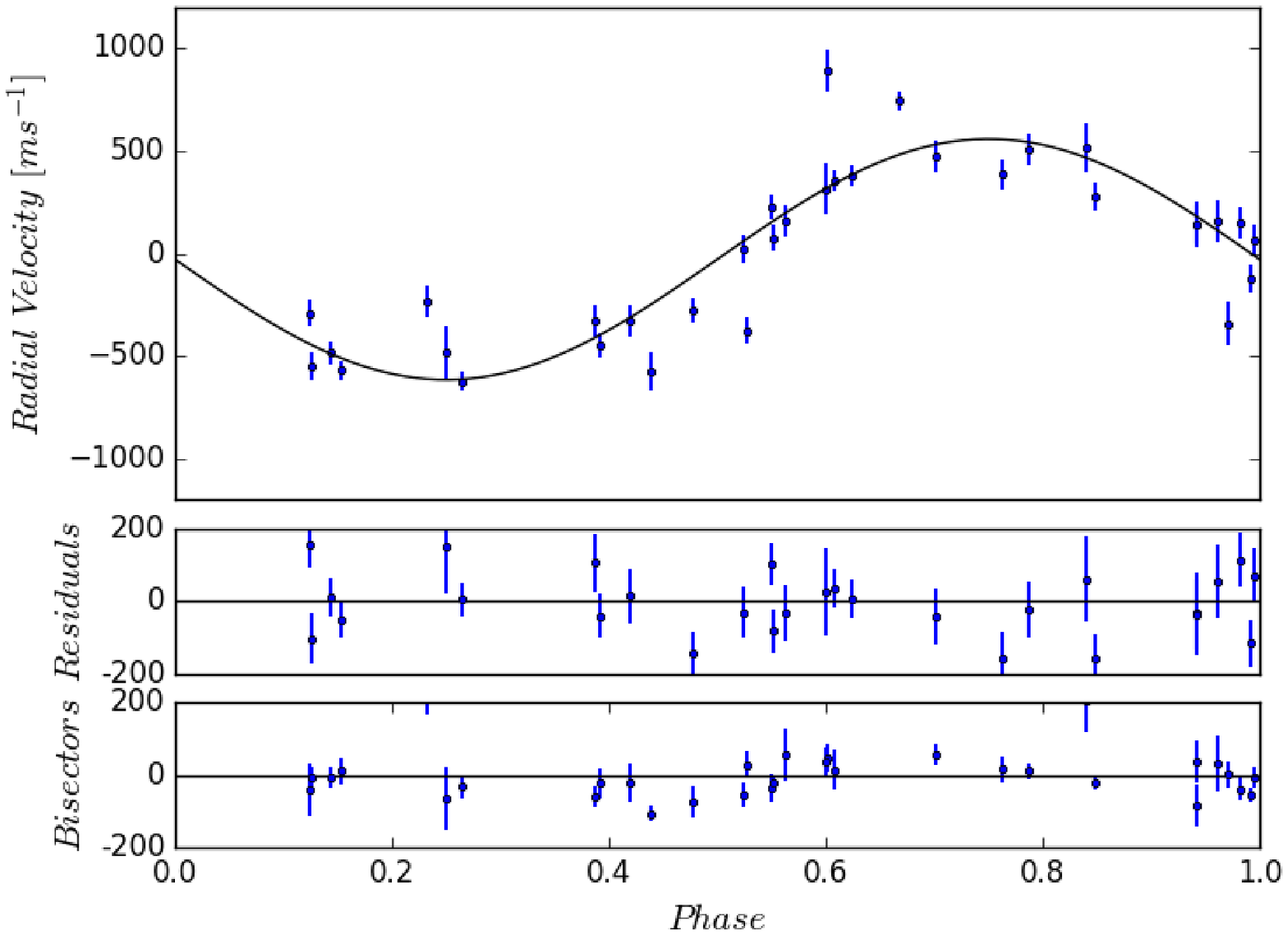}
\caption{Orbital solution for Qatar-3b, showing the velocity curve and observed velocities.}
\label{figureQ3}
\end{figure}

\begin{figure}
\centering
\includegraphics[width=8.5cm]{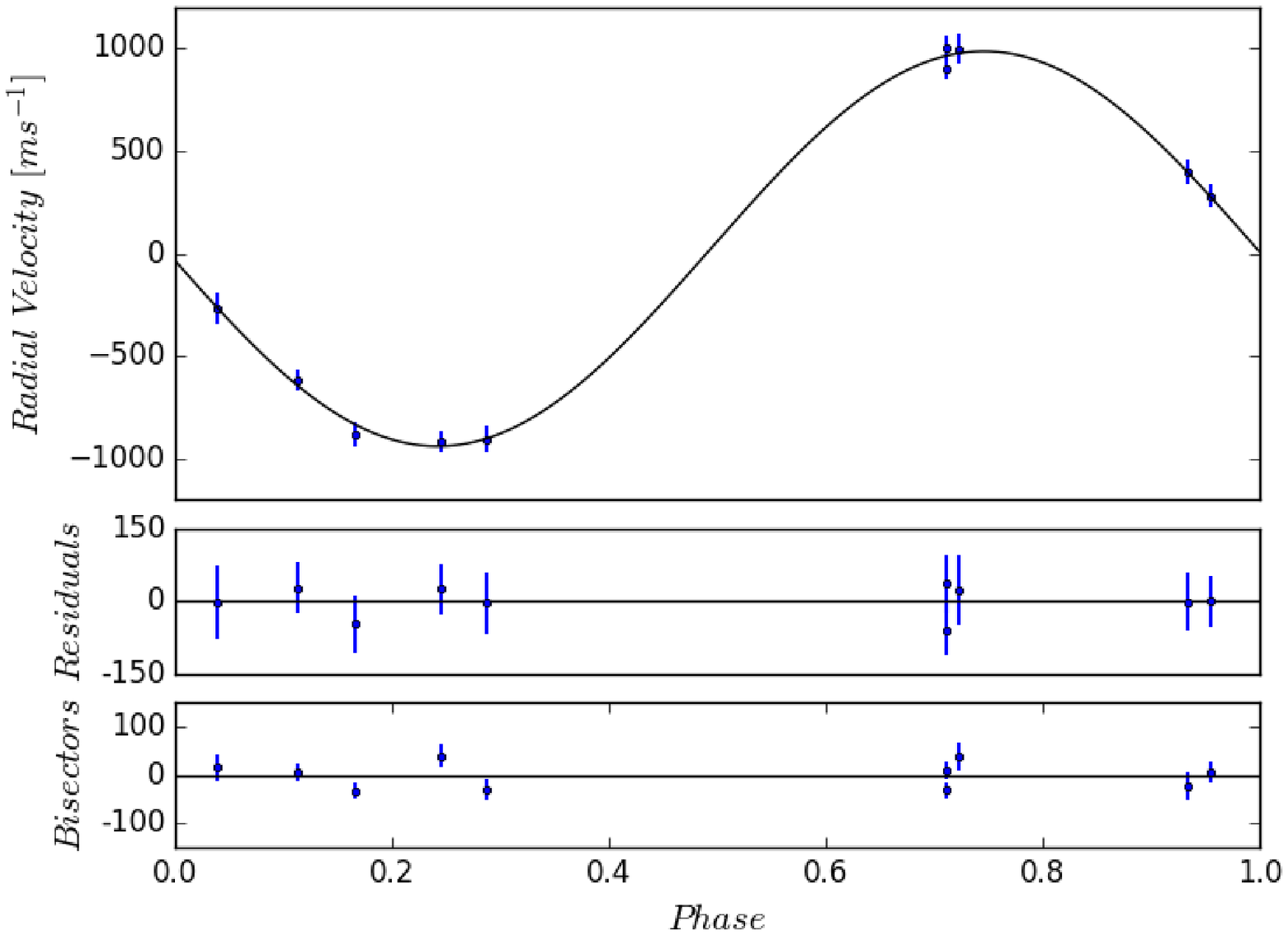}
\caption{Orbital solution for Qatar-4b, showing the velocity curve and observed velocities.}
\label{figureQ4}
\end{figure}

\begin{figure}
\centering
\includegraphics[width=8.5cm]{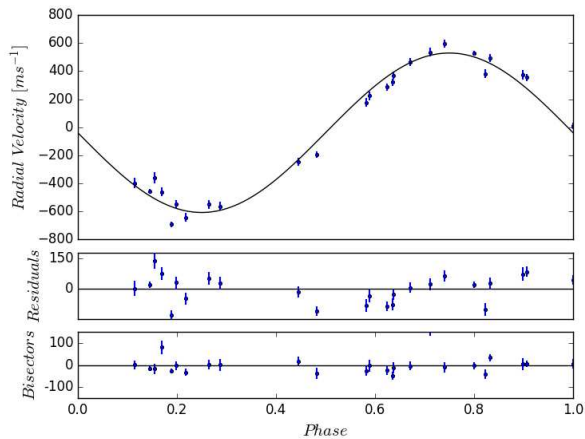}
\caption{Orbital solution for Qatar-5b, showing the velocity curve and observed velocities.}
\label{figureQ5}
\end{figure}

Relative radial velocities (RV) were derived by cross-correlating each observed 
spectrum against the strongest exposure of the same star, order by order for a 
set of echelle orders selected to have good SNRe and minimal contamination 
by telluric lines introduced by the Earth's atmosphere. These RVs are reported 
in Tables \ref{tableRVs3}, \ref{tableRVs4}, and \ref{tableRVs5} and the time units 
are in Barycentric Julian Date in Barycentric Dynamical time (BJD$_{TDB}$). The 
observation that was used for the template spectrum for each star has, by definition, 
an RV of 0.00 \kms. We define the error on the template RV as the smallest error 
of all the other errors. We also derived values for the line profile bisector spans 
(BS, lower panel in Figures \ref{figureQ3}, \ref{figureQ4}, and \ref{figureQ5}), to 
check for astrophysical phenomena other than orbital motion that might produce 
a periodic signal in the RVs with the same period as the photometric ephemerides 
for the transits.  The procedures used to determine RVs and BSs are outlined in 
\citet{buchhave2010}.

\begin{table}
\centering
\caption{Relative RVs and BS variations for Qatar-3.}
\label{tableRVs3}
\begin{tabular}{ccc}
\hline
BJD$_{TDB}$              &RV (\ms)           &BS (\ms)         \\
\hline
$ 2457233.90605 $ & $ 1259 \pm   52 $ & $ 196 \pm 28 $ \\
$ 2457237.82764 $ & $   290 \pm   75 $ & $ 157 \pm 40 $ \\
$ 2457263.83347 $ & $ 1421 \pm 101 $ & $   36 \pm 23 $ \\
$ 2457271.95553 $ & $ 1048 \pm 118 $ & $ 124 \pm 45 $ \\
$ 2457273.88262 $ & $   881 \pm   53 $ & $     5 \pm 37 $ \\
$ 2457284.80178 $ & $   694 \pm 102 $ & $   49 \pm 49 $ \\
$ 2457285.94827 $ & $   207 \pm  75 $ & $      7 \pm 26 $ \\
$ 2457288.78560 $ & $   765 \pm  57 $ & $   -14 \pm 28 $ \\
$ 2457289.87052 $ & $   684 \pm  78 $ & $   -34 \pm 20 $ \\
$ 2457291.67209 $ & $ 1012 \pm  75 $ & $    41 \pm 20 $ \\
$ 2457292.78291 $ & $     52 \pm  54 $ & $     -3 \pm 21 $ \\
$ 2457293.74375 $ & $   159 \pm  64 $ & $    21 \pm 28 $ \\
$ 2457294.85495 $ & $   202 \pm 106 $ & $     5 \pm 27 $ \\
$ 2457295.91238 $ & $     87 \pm  59 $ & $      5 \pm 26 $ \\
$ 2457296.84320 $ & $   929 \pm  70 $ & $    17 \pm 27 $ \\
$ 2457297.75075 $ & $    -13 \pm  68 $ & $     -3 \pm 22 $ \\
$ 2457298.82032 $ & $   613 \pm  62 $ & $  -18 \pm 14 $ \\
$ 2457299.79813 $ & $   685 \pm 112 $ & $   20 \pm 33 $ \\
$ 2457303.86316 $ & $   700 \pm  74 $ & $    40 \pm 45 $ \\
$ 2457304.81587 $ & $   676 \pm  94 $ & $   -61 \pm 40 $ \\
$ 2457315.61954 $ & $     61 \pm 128 $ & $  -41 \pm 62 $ \\
$ 2457318.69728 $ & $   271 \pm  59 $ & $   -42 \pm 26 $ \\
$ 2457328.84611 $ & $   576 \pm  69 $ & $   -29 \pm 27 $ \\
$ 2457332.85799 $ & $   262 \pm  62 $ & $   -21 \pm 48 $ \\
$ 2457351.66999 $ & $   947 \pm  55 $ & $ -174 \pm 22 $ \\
$ 2457357.61418 $ & $   630 \pm  75 $ & $     -4 \pm 22 $ \\
$ 2457390.61291 $ & $       0 \pm  74 $ & $      9 \pm 23 $ \\
$ 2457679.73930 $ & $      -4 \pm  91 $ & $ -115 \pm 21 $ \\
$ 2457680.76787 $ & $   847 \pm  69 $ & $   -31 \pm 16 $ \\
$ 2457681.81077 $ & $    -52 \pm  46 $ & $   -40 \pm 31 $ \\
$ 2457682.65566 $ & $   886 \pm 122 $ & $   28 \pm 38 $ \\
$ 2457683.63571 $ & $   449 \pm  66 $ & $   -63 \pm 21 $ \\
$ 2457684.62585 $ & $   243 \pm  78 $ & $   -70 \pm 29 $ \\
$ 2457685.63237 $ & $ 1078 \pm  77 $ & $      3 \pm 21 $ \\
\hline
\end{tabular}
\end{table}

\begin{table}
\centering
\caption{Relative RVs and BS variations for Qatar-4.}
\label{tableRVs4}
\begin{tabular}{ccc}
\hline
BJD$_{TDB}$              &RV (\ms)         &BS (\ms)        \\
\hline
$ 2457288.86074 $ & $  1879 \pm 73 $ & $   16 \pm 27 $ \\
$ 2457296.88246 $ & $        0 \pm 59 $ & $     6 \pm 18 $ \\
$ 2457297.86558 $ & $  1785 \pm 52 $ & $  -32 \pm 15 $ \\
$ 2457327.79409 $ & $     -23 \pm 64 $ & $   40 \pm 25 $ \\
$ 2457356.60189 $ & $     -34 \pm 52 $ & $  -30 \pm 22 $ \\
$ 2457390.66755 $ & $    263 \pm 52 $ & $  -31 \pm 17 $ \\
$ 2457417.61546 $ & $    613 \pm 76 $ & $   10 \pm 17 $ \\
$ 2457653.93140 $ & $  1274 \pm 60 $ & $   40 \pm 29 $ \\
$ 2457749.65461 $ & $  1162 \pm 52 $ & $  -23 \pm 27 $ \\
$ 2457772.68491 $ & $  1880 \pm 59 $ & $     5 \pm 21 $ \\

\hline
\end{tabular}
\end{table}

\begin{table}
\centering
\caption{Relative RVs and BS variations for Qatar-5.}
\label{tableRVs5}
\begin{tabular}{ccc}
\hline
BJD$_{TDB}$              &RV (\ms)         &BS (\ms)        \\
\hline
$ 2457292.73838 $ & $   992 \pm 29 $ & $ 160 \pm 29 $ \\
$ 2457296.93621 $ & $      -6 \pm 31 $ & $   81 \pm 31 $ \\
$ 2457298.84661 $ & $   950 \pm 29 $ & $   35 \pm 17 $ \\
$ 2457299.77524 $ & $     95 \pm 38 $ & $  -18 \pm 25 $ \\
$ 2457318.73669 $ & $ 1051 \pm 28 $ & $    -9 \pm 24 $ \\
$ 2457327.82834 $ & $   828 \pm 34 $ & $     5 \pm 28 $ \\
$ 2457328.69019 $ & $    -92 \pm 29 $ & $    -3 \pm 19 $ \\
$ 2457329.81549 $ & $   680 \pm 32 $ & $    -3 \pm 27 $ \\
$ 2457332.83362 $ & $   824 \pm 28 $ & $  -12 \pm 25 $ \\
$ 2457345.72455 $ & $     58 \pm 37 $ & $     2 \pm 21 $ \\
$ 2457346.67754 $ & $   210 \pm 28 $ & $   16 \pm 22 $ \\
$ 2457347.69820 $ & $   985 \pm 16 $ & $    -3 \pm 16 $ \\
$ 2457348.69169 $ & $       0 \pm 16 $ & $  -17 \pm 10 $ \\
$ 2457349.66214 $ & $   261 \pm 23 $ & $  -39 \pm 25 $ \\
$ 2457350.64354 $ & $   838 \pm 31 $ & $  -41 \pm 22 $ \\
$ 2457351.69765 $ & $  -235 \pm 21 $ & $  -29 \pm 12 $ \\
$ 2457354.66154 $ & $  -186 \pm 31 $ & $  -34 \pm 17 $ \\
$ 2457355.71215 $ & $   632 \pm 30 $ & $  -28 \pm 22 $ \\
$ 2457356.64588 $ & $   812 \pm 27 $ & $     5 \pm 16 $ \\
$ 2457357.67226 $ & $    -95 \pm 31 $ & $     2 \pm 21 $ \\
$ 2457358.71044 $ & $   741 \pm 25 $ & $  -25 \pm 21 $ \\
$ 2457360.61730 $ & $  -109 \pm 32 $ & $     1 \pm 28 $ \\
$ 2457361.62125 $ & $   775 \pm 28 $ & $  -49 \pm 19 $ \\
$ 2457362.67051 $ & $   464 \pm 25 $ & $     6 \pm 22 $ \\
$ 2457364.60306 $ & $   919 \pm 27 $ & $    -4 \pm 20 $ \\
\hline
\end{tabular}
\end{table}

To illustrate the quality of the orbital solutions provided by our relative radial 
velocities, we fit circular orbits with the epoch and period set to the final 
ephemerides values from the global analysis. The key parameters for these 
orbital solutions are reported in Table \ref{tableOrbPar}, and the corresponding 
radial velocity curves and individual observations are plotted in Figures \ref{figureQ3},
\ref{figureQ4}, and \ref{figureQ5}.  Note that the relative gamma velocity is the 
center-of-mass velocity using the relative velocities. 

The values of the correlation coefficient between Bisectors and RVs for Qatar-3b
(0.313), Qatar-4b (0.020) and Qatar-5b (0.153) are low and suggest the 
correlation is not significant in all three cases. We do not calculate the FWHM of 
the correlation function. As an alternative approach, we run SPC and derive the 
\vsini\ values as a measure of the broadening.

To get the absolute gamma (center-of-mass) velocity for a system where we 
use the multi-order relative velocities to derive the orbital solution, we have to 
provide an absolute velocity for the observation that was used for the template 
when deriving the relative velocities.  By definition that observation is assigned 
a relative velocity of 0.00 \kms. To derive an absolute velocity for that observation, 
we correlate the Mg b order against the template from the CfA library of synthetic 
templates that gives the highest peak correlation value.  Then we add the 
relative gamma velocity from the orbital solution, and also correct by $-0.61$ \kms, 
mostly because the CfA library does not include the gravitational redshift.  This 
offset has been determined empirically by many observations of IAU Radial 
Velocity Standard Stars. We quote an uncertainty in the resulting absolute 
velocity of $\pm 0.1$ \kms, which is an estimate of the residual systematic 
errors in the IAU Radial Velocity Standard Star system.

\begin{table}
\centering
\caption{Initial Orbital Parameters.}
\label{tableOrbPar}
\begin{tabular}{lccc}
\hline
Orbital Parameter        & Qatar-3b          & Qatar-4b           & Qatar-5b            \\
\hline
Semi-amplitude $K$ (\ms)   & $ 594 \pm 70    $ & $ 1087 \pm 88    $ & $ 570 \pm 17      $ \\
Relative $\gamma$  (\ms)   & $ 542 \pm 44    $ & $  927  \pm 74    $ & $ 416 \pm 13      $ \\
Absolute $\gamma$ (\kms) & $ +6.04 \pm 0.1 $ & $ -28.76 \pm 0.1 $ & $ -9.54 \pm 0.1   $ \\
RMS RV residuals   (\ms)    & $ 217           $ & $ 263            $ & $ 67              $ \\
Number of RVs                    & $ 34            $ & $ 10              $ & $ 25              $ \\
Reduced $\chi^{2}$ RV & 7.1& 6.9& 4.5 \\    
\hline
\end{tabular}
\end{table}


\subsection{Follow-up photometry} \label{subsec:folupphot}

\begin{figure}
\centering
\includegraphics[width=8.5cm]{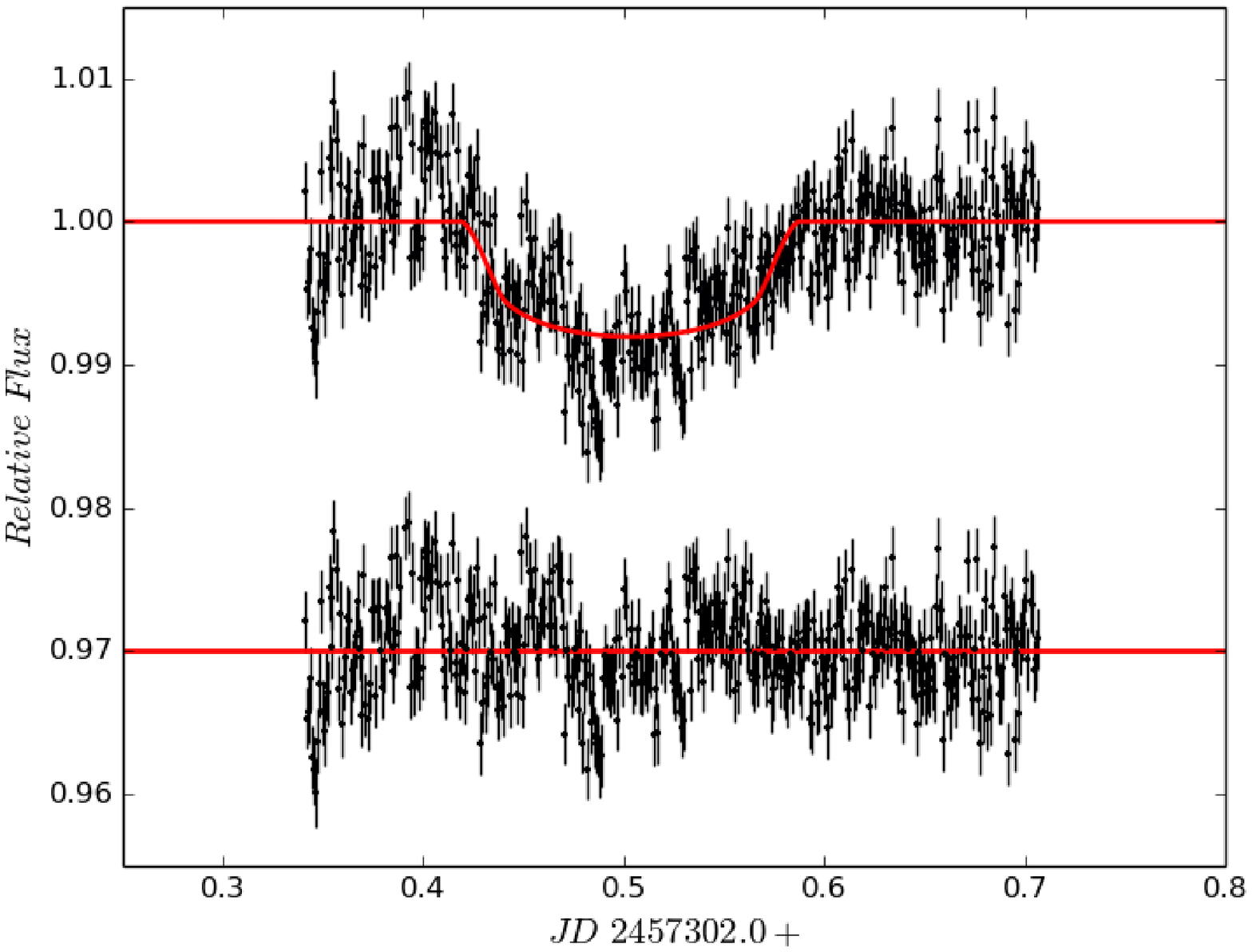} \\
\includegraphics[width=8.5cm]{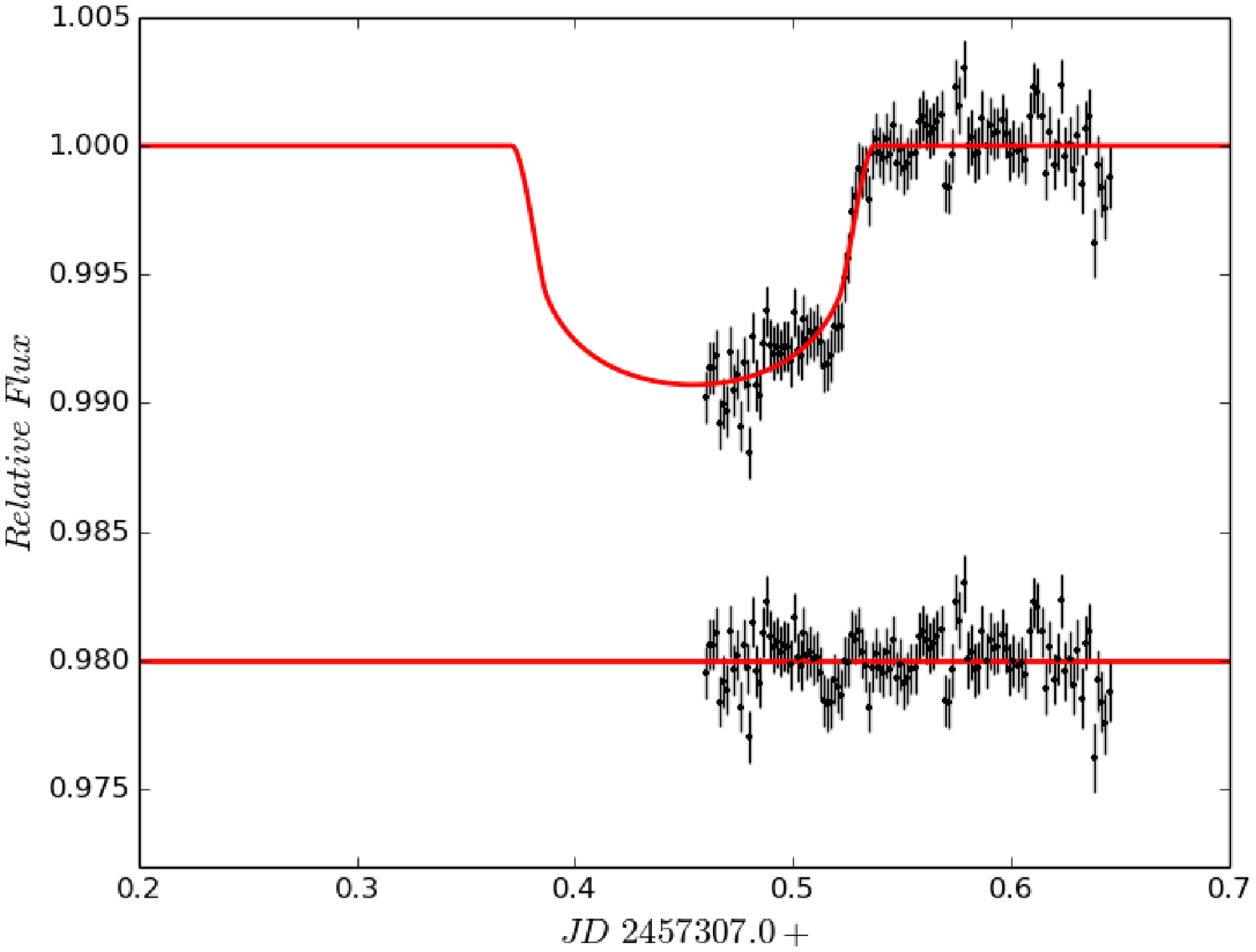}
\caption{I-band follow-up light curves of Qatar-3b, obtained on 06/10/2015 (top panel) 
and 11/10/2015 (bottom panel) using the 1.23m Zeiss telescope at the Calar Alto 
observatory. The best-fit transit model overlayed in red (see text for details). The light 
curves suffer by some extra noise due to the poor weather conditions.}
\label{fig2}
\end{figure}

\begin{figure}
\centering
\includegraphics[width=8.5cm]{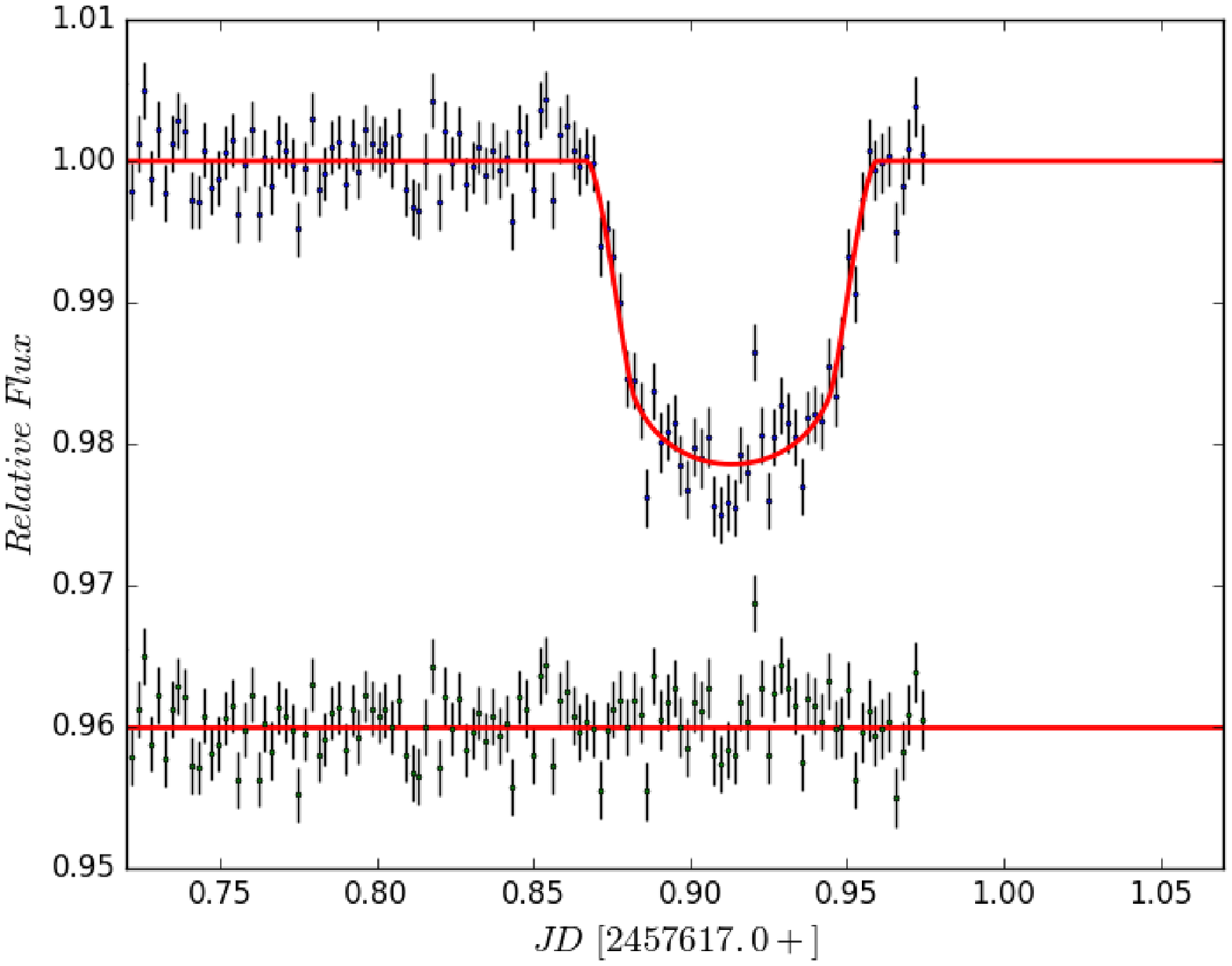} \\ 	
\includegraphics[width=8.5cm]{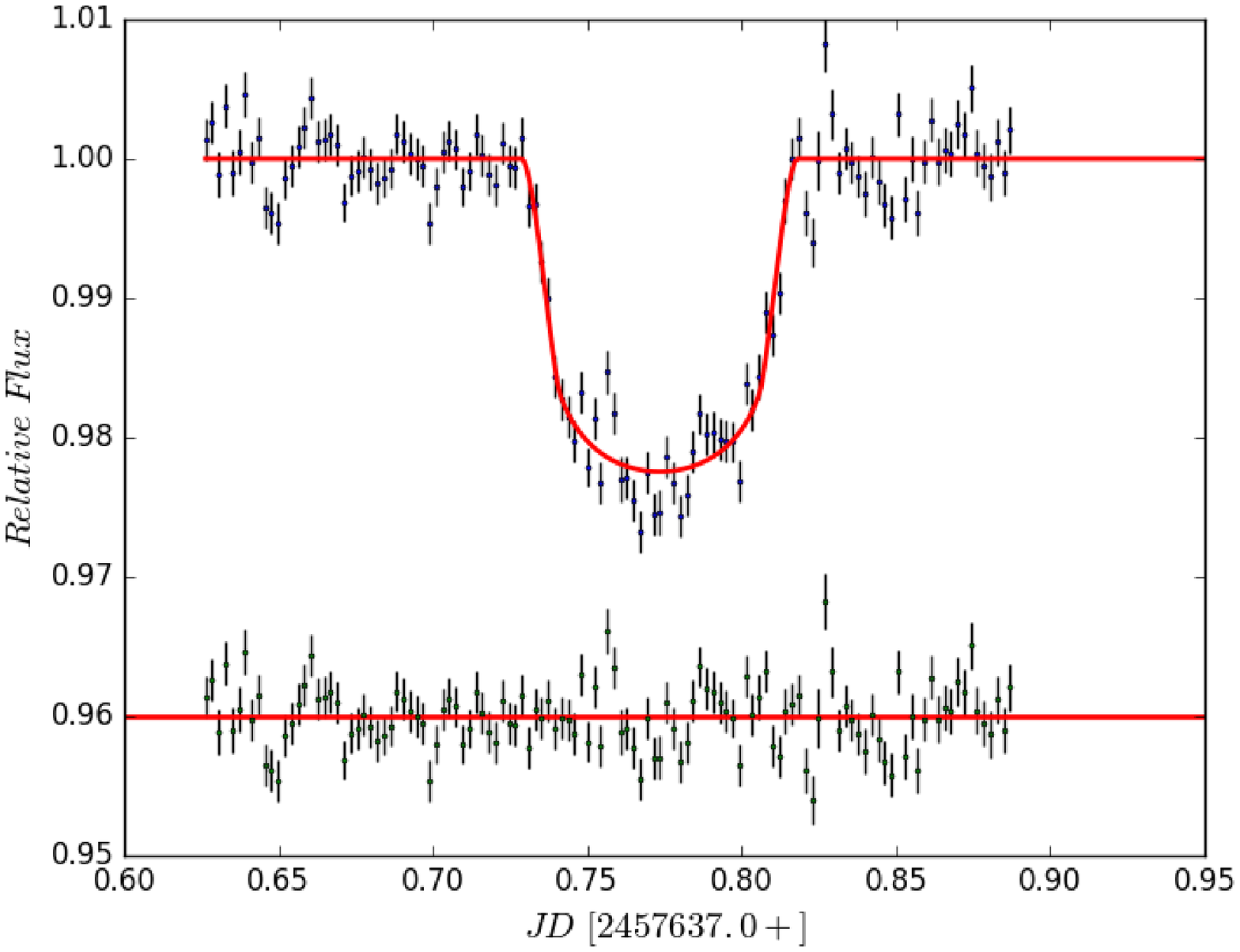} \\ 	
\includegraphics[width=8.5cm]{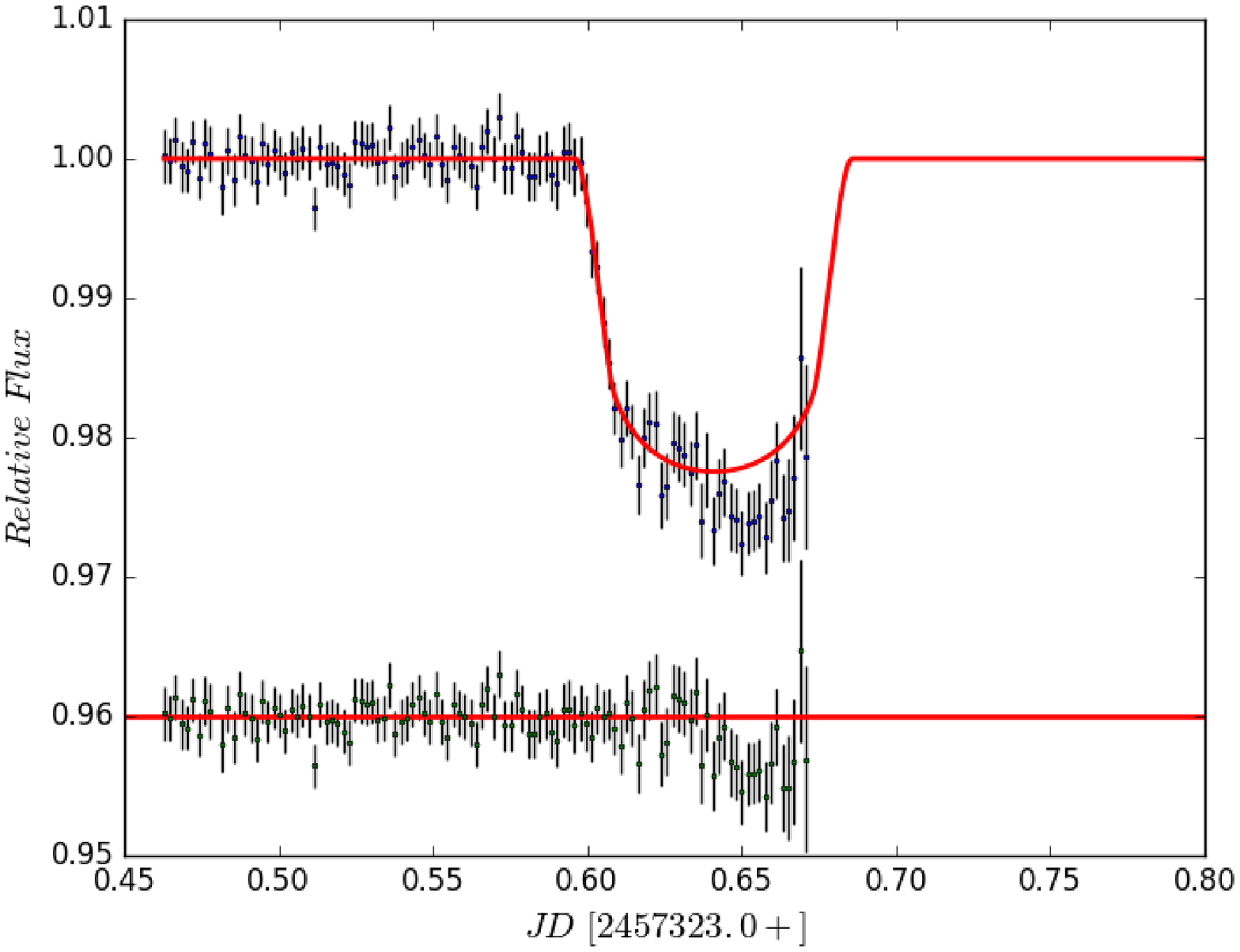} 
\caption{Same as Fig.\,\ref{fig2}, but for Qatar-4b, observed on 27/10/2015 using 
the 1.23m Zeiss telescope at the Calar Alto observatory, and on 17/08/2016 and 
06/09/2016 using the QFT. }
\label{fig3}
\end{figure}

\begin{figure}
\centering
\includegraphics[width=8.5cm]{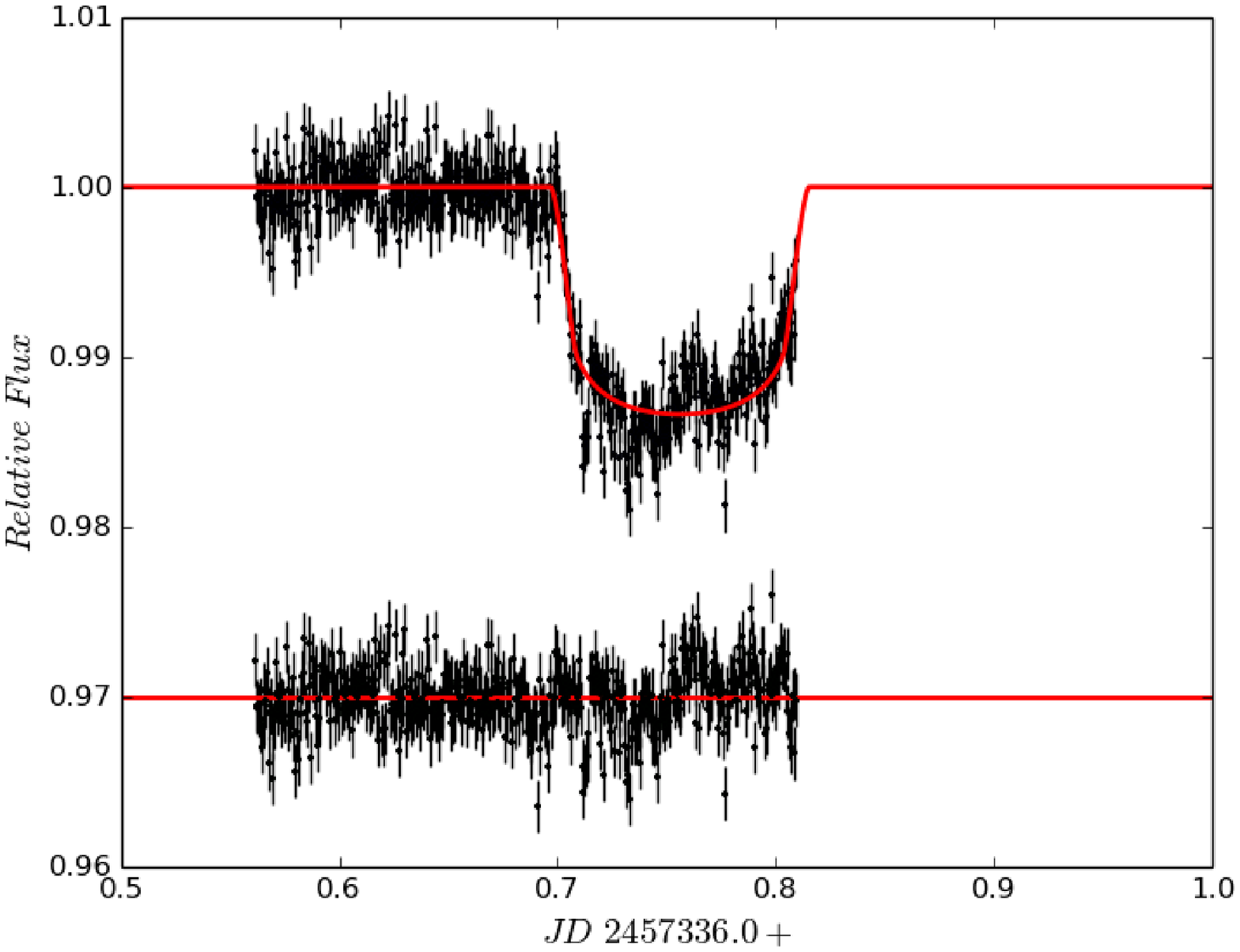}	 
\caption{Same as Fig.\,\ref{fig2}, but for Qatar-5b, observed on 10/11/2015 with KeplerCam. }
\label{fig4}
\end{figure}

Follow-up photometric observations for Qatar-3b and Qatar-4b were obtained with the 
1.23\,m Zeiss Telescope at the Calar Alto Observatory (CAHA, Spain), using a Cousins-I 
filter and an exposure time of 60s per frame. For all observations, the telescope was 
defocused and data reduction was carried out using the \texttt{DEFOT} pipeline 
\citep{south1, south2}. Qatar-3b was observed on two occasions, on the $6^{th}$ and the 
$11^{th}$ of October 2015, while a half-transit of Qatar-4b was observed on the $27^{th}$ 
of October 2015. Two additional transits of Qatar-4b were obseved with the 50 cm QES 
Follow-up Telescop (QFT) installed at the QES station hosted by the New Mexico Skies 
Observatory on the nights of $17^{th}$ August and $6^{th}$ September 2016. QFT 
is equipped with Andor iKon-M 934 deep depletion, back illuminated CCD camera 
optimized for follow-up photometric studies. Qatar-4b light curves were obtained through 
a Johnson-I filter using defocusing technique similar to our Calar Alto observations and 
an exposure time of 180 s per frame.  
Follow-up light curve for Qatar-5b was obtained using the KeplerCam 
on the 1.2m telescope at the Fred L.\, Whipple Observatory on Mount Hopkins, Arizona 
on the night of $10^{th}$ November 2015. KeplerCam is equipped with a single 
4K $\times$ 4K CCD covering an area of 23$^{\prime} \times 23^{\prime}$ on the sky. 
The observations were obtained through a SDSS-$i^{\prime}$ filter. 
Figures\,\ref{fig2},\,\ref{fig3}\,\&\,\ref{fig4} show the follow-up light curves together with
the model fits described in Section \ref{subsec:EXOFAST}.

To better determine the transiting systems ephemerides we fit the follow-up photometric 
curves with a transiting model following the prescription outlined in \cite{pal}. In short, 
the \cite{pal} method uses analytical expressions to evaluate the partial derivatives of
the flux decrease function for an eclipsed star, under the assumption of quadratic limb 
darkening. \cite{pal} equations allow for a clear separation between terms depending 
only on the limb darkening coefficients, and terms depending only on the occultation 
geometry.

After the model fit, we estimate the $T_{C}$ and calculate the best ephemerides. For 
the current ephemerides, we used the $T_{C}$ from the best model fit of the light curve. 
Ephemerides are listed in Table \ref{ExoQ345}. Note that we follow the standard 
procedure and we  did not include the discovery light curves in the physical parameter 
analysis. The discovery light curve data points have too large errorbars, and we used 
only the follow-up high precision light curves, in order to reduce the errors in the 
physical parameters. 

\section{Analysis and Results} \label{sec:anares} 

\subsection{Stellar Parameters} \label{subsec:SPC}

To improve the characterization of the three host stars, we analyzed the TRES spectra 
using the Stellar Parameter Classification (SPC) tool developed by \cite{buchhave2012}.  
In brief, the SPC cross correlates the observed spectrum with a library of synthetic spectra 
from Kurucz model atmospheres and finds the stellar parameters from a multi-dimensional 
surface fit to the peak correlation values. We used the ATLAS9 grid of models with the new 
Opacity Distribution Functions from \cite{ODF}. In addition, the stellar parameters --- effective 
temperature ($T_{\rm eff}$), metallicity ([m/H]), surface gravity (\logg), and projected 
rotational velocity $v \sin{i}$ --- for the hosts were derived from the co-added spectra of 
each star through spectral modelling using the Spectroscopy Made Easy (SME) package 
(\cite{sme}). 

We note that the values of \teff\,, estimated via SPC and SME, are within 3.5$\sigma$ 
(for Qatar-3), 2$\sigma$ (Qatar-4), and 1$\sigma$ (Qatar-5) of each other, while the 
\logg\, and $v \sin{i}$ values for all three stars are essentially the same (differences are 
less than 1$\sigma$). The only noticeable difference is metallicity ([m/H]), where SME 
gives systematically lower values relative to SPC by 0.3. We examine the effect of these 
differences on the calculated planetary parameters (mass and radius) in the next section.

In Table 6 we also provide the ages of the host stars using gyrochronology equations from 
\cite{brown} -- eq.\ 1, assuming that the stellar rotation axis is perpendicular to the orbital 
plane. We found that the ages for all three stars are $\tau_{\rm gyr, Q3}=0.31$ Gyr, 
$\tau_{\rm gyr, Q4}=0.17$ Gyr and $\tau_{\rm gyr, Q5}=0.53$ Gyr for Qatar-3b, Qatar-4b and 
Qatar-5b, respectively. Additionally, using model isochrones from \cite{dotter} and the input 
parameters for Table 6, we calculate independent values for the ages of the host stars 
$0.1 < \tau_{\rm iso,Q3,Q4,Q5} < 0.3$ Gyr. All host stars are relatively young stars, which is 
basically consistent with their relatively fast rotation ($\vsini > 5$ \kms). We note that 
previous studies (\cite{maxted}, \cite{brown}) show that in general gyrochronology suggests 
younger age than isochrone models. In our case the ages from both 
methods---gyrochronology and model isochrones---are generally consistent with each other. 


\subsection{Planetary System Parameters} \label{subsec:EXOFAST}

To determine the physical parameters of the three planetary systems we run a global
solution of the available RV and transit photometric data using the EXOFAST package 
(\cite{exofast}). The transit light curves include only the follow-up photometric 
data and not the discovery light curves. As described by the authors, the EXOFAST 
performs a simultaneous fit of the RV and/or transit data for a single planet. In our case, 
for all the systems, we fixed the planetary orbital period to the value determined from the 
transits ephemerides and set the initial stellar parameters (\teff, \logg, \feh) to the values 
determined from the spectroscopic analysis of the host stars.

To quantify the effect different sets of values for \teff, \logg, \feh, estimated via SPC and 
SME, have on the calculated values for the planetary mass and radius ($M_{P}$, $R_{P}$), 
we fed EXOFAST with the sets of initial stellar parameters determined by SPC and SME 
separately and compared the results. We remind the reader, that internally EXOFAST 
uses the Torres relations (\cite{torres})--- calibrations based on accurate ($\leq$3\%) 
masses and radii from detached binary systems --- to determine the masses and radii 
of the host stars. These relations are valid for main sequence stars above 0.6 \msun, 
and we note that all our stars have estimated ages and masses well within the range 
covered by the Torres relations.

The nature of the Torres relations is such, that $M_{*}$ and $R_{*}$ are only weakly 
dependent on metallicity. As a result, the SPC and SME sets of values lead to 
very similar host stellar masses and radii --- indistinguishable for Qatar-3; 4\%, and 
5\% difference in $M_{*}$ and $R_{*}$, respectively for Qatar-4; and 4\%, and 2\% 
difference in $M_{*}$ and $R_{*}$, respectively for Qatar-5. The biggest differences 
are in the luminosity of the host stars where the SME sets of parameters lead to a 13\% 
more luminous star for Qatar-3, and 15\% more luminous star for Qatar-4, but essentially 
the same for Qatar-5. 

Most importantly differences in stellar parameters produced via SPC and SME lead to 
insignificant differences in the derived values for the planetary masses and radii. In all 
three cases --- Qatar-3b, Qatar-4b, and Qatar-5b --- these are well within 1$\sigma$ of 
the uncertainty. For this reason, in all tables, we list only the values derived with the 
initial set of stellar parameters determined via SPC.  

The initial evaluation of the fits to the RV curves indicated they were all well described 
by circular orbits, i.e., $e = 0$. On one hand this is not surprising, as all three planets 
have short period orbits that are expected to have circularized. In addition, in the case 
of Qatar-4b, the RV curve has relatively few points and does not warrant a detailed 
search for an eccentric solution. In the case of Qatar-3b and Qatar-5b, we searched for 
eccentric solutions as well, but in both cases the results were essentially indistinguishable 
from $e = 0$ at the $\le2\sigma$ level. Consequently, in our global fits we kept the 
eccentricity fixed at $e = 0$. In addition, the period of each planet was kept fixed at 
the value determined by the transit ephemerides by in practice allowing it to vary only 
at the insignificant $10^{-5}$ d level. 

Table \ref{ExoQ345} summarises the physical parameters of the planets. The best fit for 
both radial velocity and photometric light curves is coming from EXOFAST. The Safronov 
numbers for each planet are not used in the current paper and are provided in Table 
6 for completeness, as they may be useful for other studies.

\begin{table*}
\caption{Median values and 68\% confidence intervals. We assume $R_{\odot}$=696342.0\,km, 
$M_{\odot}$=1.98855$\times 10^{30}$\,kg, $R_{\rm J}$ = 69911.0\,km, 
$M_{\rm J}$=1.8986$\times 10^{27}$\,kg and 1 AU=149597870.7 km.}
\label{ExoQ345}
\begin{tabular}{lcccc}
\hline
Parameter & Units & Qatar-3b & Qatar-4b & Qatar-5b\\
\hline
Stellar Parameters: & & & & \\
    ~~~$M_{*}$\dotfill     &Mass (\msun)\dotfill  & $1.145\pm0.064$ & $0.896\pm0.048$& $1.128\pm0.056$\\
    ~~~$R_{*}$\dotfill      &Radius (\rsun)\dotfill & $1.272\pm0.14$   & $0.849\pm0.063$& $1.076\pm0.051$\\
    ~~~$L_{*}$\dotfill       &Luminosity (\lsun)\dotfill & $1.90\pm0.46$ & $0.481\pm0.076$& $1.138\pm0.12$\\
    ~~~$\rho_*$\dotfill     &Density (g/cm$^{3}$)\dotfill & $0.78\pm0.20$  & $2.07\pm0.038$& $1.286\pm0.15$\\
    ~~~$\log(g_*)$\dotfill &Surface gravity (cgs)\dotfill & $4.286\pm0.079$& $4.533\pm0.058$& $4.427\pm0.035$\\
    ~~~$\teff$\dotfill        &Effective temperature (K)\dotfill & $6007\pm52$    & $5215\pm50$  & $5747\pm49$     \\
    ~~~$\feh$\dotfill        &Metallicity\dotfill & $-0.041\pm0.081$  & $0.102\pm0.079$  & $0.377\pm0.080$   \\
    ~~~$age$\dotfill        &Age [Gyr]\dotfill & $0.310\pm0.001$  & $0.170\pm0.010$  & $0.530\pm0.004$   \\
    ~~~$P_{rot}$\dotfill   &Rotation period [days]\dotfill & $6.31$  & $6.05$  & $12.10$   \\
Planetary Parameters: & & & & \\
    ~~~$P$\dotfill   &Period (days)\dotfill & $2.5079204$ & $1.8053564$ & $2.8792319$\\
    ~~~$a$\dotfill   &Semi-major axis (AU)\dotfill & $0.03783\pm0.00069$& $0.02803\pm0.00048$& $0.04127\pm0.00067$\\
    ~~~$M_{P}$\dotfill &Mass (\mj)\dotfill & $4.31\pm0.47$& $5.36\pm0.20$& $4.32\pm0.18$\\
    ~~~$R_{P}$\dotfill &Radius (\rj)\dotfill & $1.096\pm0.14$& $1.135\pm0.11$& $1.107\pm0.064$\\
    ~~~$\rho_{P}$\dotfill &Density (g/cm$^{3}$)\dotfill & $4.0\pm1.2$& $4.50\pm1.00$& $3.95\pm0.58$\\
    ~~~$\log(g_{P})$\dotfill &Surface gravity\dotfill & $3.942\pm0.10$& $4.010\pm0.078$& $3.940\pm0.044$\\
    ~~~$T_{eq}$\dotfill &Equilibrium Temperature (K)\dotfill & $1681\pm84$& $1385\pm50$& $1415\pm31$\\
    ~~~$\Theta$\dotfill &Safronov Number\dotfill & $0.256\pm0.035$& $0.293\pm0.025$& $0.284\pm0.016$\\
    ~~~$\fave$\dotfill &Incident flux (\fluxcgs)\dotfill & $1.81\pm0.39$& $0.835\pm0.12$& $0.910\pm0.082$\\
    ~~~$e$\dotfill &Eccentricity (fixed)\dotfill & $0$& $0$ & $0$\\
RV Parameters: & & & & \\
    ~~~$K$\dotfill &RV semi-amplitude (m/s)\dotfill & $587\pm58$& $957\pm16$& $568\pm15$\\
    ~~~$\gamma$\dotfill &Systemic velocity (m/s)\dotfill & $542\pm36$& $921\pm13$& $416\pm11$\\
Primary Transit Parameters: & & & & \\
    ~~~$T_C$\dotfill &Time of transit (\bjdtdb)\dotfill & $2457302.453004\pm0.00010$& $2457637.77370\pm0.00046$& $2457336.758242\pm0.000098$\\
    ~~~$R_{P}/R_{*}$\dotfill &Radius of planet in stellar radii\dotfill & $0.0888\pm0.0018$& $0.1379\pm0.0030$& $0.1061\pm0.0013$\\
    ~~~$a/R_{*}$\dotfill &Semi-major axis in stellar radii\dotfill & $6.39\pm0.59$& $7.09\pm0.48$& $8.27\pm0.34$\\
    ~~~$u_1$\dotfill &linear limb-darkening coeff\dotfill & $0.258\pm0.047$& $0.400\pm0.048$& $0.323\pm0.043$\\
    ~~~$u_2$\dotfill &quadratic limb-darkening coeff\dotfill & $0.293\pm0.050$& $0.233\pm0.050$& $0.271\pm0.048$\\
    ~~~$i$\dotfill &Inclination (degrees)\dotfill & $86.8\pm2.0$& $87.5\pm1.6$& $88.74\pm0.87$\\
    ~~~$b$\dotfill &Impact Parameter\dotfill & $0.35\pm0.22$& $0.31\pm0.19$& $0.18\pm0.16$\\
    ~~~$\delta$\dotfill &Transit depth\dotfill & $0.00789\pm0.00032$& $0.01902\pm0.00083$& $0.01126\pm0.00028$\\
    ~~~$\tau$\dotfill &Ingress/egress duration (days)\dotfill & $0.0120\pm0.0028$& $0.0118\pm0.0022$& $0.01201\pm0.0013$\\
    ~~~$T_{14}$\dotfill &Total duration (days)\dotfill & $0.1296\pm0.0034$& $0.0891\pm0.0019$& $0.1212\pm0.0015$\\
\hline
\end{tabular}
\end{table*}

\noindent

\section{Discussion} \label{sec:disc}

Using the equations from \cite{leconte} and \cite{jackson}, we calculate the tidal interaction 
time-scale for the eccentricity evolution of the systems. We used the values from Table 
\ref{ExoQ345} of $M_{\star}$, $R_{\star}$, $M_{P}$, $R_{P}$, assuming tidal quality factors 
of $Q_{\star}=10^{6.5}$ and $Q_{P}=10^{5.5}$. The three rotation periods given in 
Table \ref{ExoQ345} --- $P_{Q3}=6.31$ d, $P_{Q4}=6.05$ and $P_{Q5}=12.10$ d, respectively 
--- are calculated using the stellar radii from our solutions, the \vsini\ form our spectra, and 
assuming the stellar rotation axis and the planet orbit are coplaner. Finally the time-scales for 
the eccentric evolution is $\tau_{Q3}=0.133$ Gy, $\tau_{Q4}=0.0870$ Gy, $\tau_{Q3}=0.544$ 
Gy for the three systems, respectively. The eccentricity evolution timescale for Qatar-5b, is 
approximately equal with its age (Table \ref{ExoQ345}). 

\begin{figure}
\centering
\includegraphics[width=8.5cm]{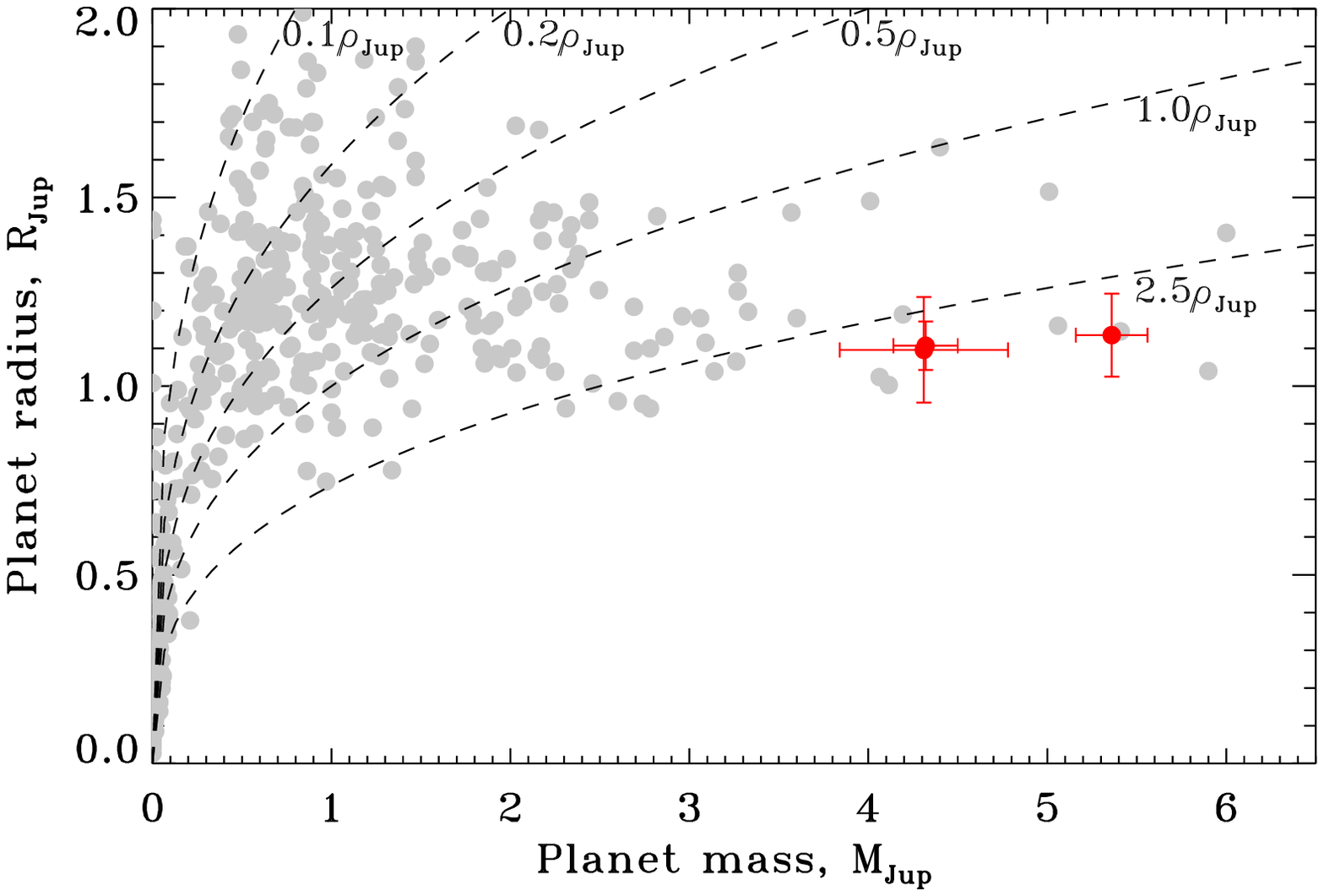} \\
\includegraphics[width=8.5cm]{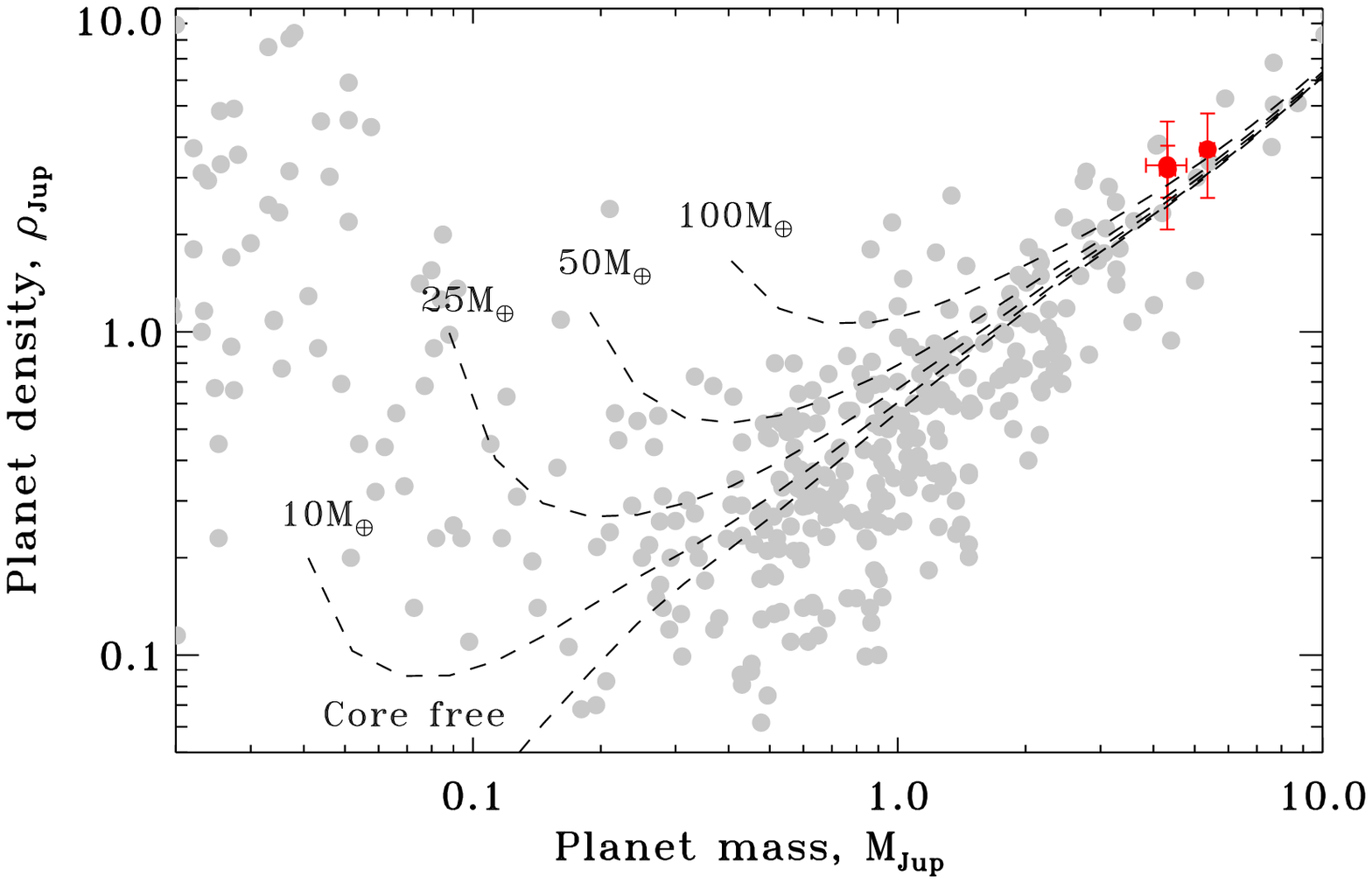} 
\caption{{\it Top:} Mass-radius diagram of known transiting exoplanets. Light gray points 
represent data for well-studied planets from the TEPcat. Error bars are suppresed for clarity.
Qatar-3b, Qatar-4b and Qatar-5b are shown as red points with error bars. Dashed lines show 
constant density for 0.1, 0.25, 0.5, 1.0, and 2.5 $\rho_{\rm Jup}$. {\it Bottom:} The mass-density
diagram for currently known exoplanets (also taken form TEPcat). As in the previous panel the 
new planets are shown as red dots with error bars. Models for giant planets with different core 
masses are drawn as dashed lines for comparison.}
\label{mass_radius_density}
\end{figure}

The three planets presented here fall in the arera of heavy hot Jupiters with masses in the 
range 4-6$M_{J}$ and densities 4-5 g\,cm$^{\rm -3}$ (see Table \ref{ExoQ345}). Their equilibrium 
temperatures place them in the pL class of planets following the \cite{F2008} nomenclature. To put 
the properties of the three new planets in perspective we show their positions on the planet 
mass-radius (Fig.\,\ref{mass_radius_density}, top) and mass-density (Fig.\,\ref{mass_radius_density}, 
bottom) diagrams and compare them with data for the well studied transiting exoplanets from 
TEPcat\footnote{The Transiting Extrasolar Planet Catalog (TEPcat) is available at 
http://www.astro.keele.uk/jkt/tepcat}. On both the mass-radius and mass-density diagrams the 
three new planets occupy the sparsely populated area of relatively heavy and dense planets on 
one end of the parameter space. On each panel of Fig.\,\ref{mass_radius_density}, Qatar-3b and 
Qatar-5b project almost on top of each other with Qatar-4b close nearby and all three are close to 
the theoretical models and in company of some other observed planets. On the mass-density 
diagram we also show the 0.3 Gyr model isochrones from \cite{F2007}\footnote{Models are 
available at http://www.ucolick.org/~jfortney/models.htm} for giant planets with different core 
mass values at a distance of 0.045 AU. The three planes occupy the area of the mass-density 
diagram which is insensitive to a particular core mass value. 

\section{Conclusions} \label{sec:Con}

Qatar-3b, Qatar-4b, and Qatar-5b are three new transiting hot Jupiters hosted by K1V, G0V, 
and G2V stars respectively. All three are short period planets ($P_{Q3b}$=2.50792, 
$P_{Q4b}$=1.80536, and $P_{Q5b}$=2.87923 days) with with masses and radii 
($M_{Q3b}$=4.31$M_{\rm J}$ $R_{Q3b}$=1.096$R_{\rm J}$, $M_{Q4b}$=5.36$M_{\rm J}$ 
$R_{Q4b}$=1.135$R_{\rm J}$, $M_{Q5b}$=4.32$M_{\rm J}$ $R_{Q5b}$=1.107$R_{\rm J}$) 
in the expected regime for hot Jupiters, and densities in the range 4-5 g/cm$^{3}$. The 
planets look similar to other members of the hot Jupiter family on the mass--radius and 
mass--density diagrams. We note, however, that all three planets reside in the sparsely 
populated heavy-mass end ($M > 4$\mj) on the mass-radius diagram. Future follow-up 
observations will help characterize these planets in greater detail and help shed light on 
some of the peculiarities.

\noindent

\section*{Acknowledgements}

This publication is supported by NPRP grant no. X-019-1-006 from the Qatar National 
Research Fund (a member of Qatar Foundation). The statements made herein are solely 
the responsibility of the authors. D.F.E. is funded by the UK Science and Technology 
Facilities Council.




\begin{thebibliography}{}

\bibitem[\protect\citeauthoryear{Alonso et al.}{2004}]{tres} Alonso R. et al, 2004, ApJ, 613L, 153
\bibitem[\protect\citeauthoryear{Alsubai et al.}{2013}]{alsubai} Alsubai K. et al. 2013, Acta Astron., 63, 465
\bibitem[\protect\citeauthoryear{Barnes}{2007}]{barnes} Barnes S., 2007, ApJ, 669, 1167
\bibitem[\protect\citeauthoryear{Bakos et al.}{2004}]{hatn} Bakos G. et al., 2004, PASP, 116, 266
\bibitem[\protect\citeauthoryear{Batalha et al.}{2013}]{batalha} Batalha N. et al. 2013, ApJS, 204, 24
\bibitem[\protect\citeauthoryear{Bramich}{2008}]{dbdia} Bramich D. M. 2008, MNRAS, 386, L77
\bibitem[\protect\citeauthoryear{Brown}{2014}]{brown} Brown D. J. A. 2014, MNRAS, 442, 1844
\bibitem[\protect\citeauthoryear{Buchhave et al.}{2010}]{buchhave2010} Buchhave L. A. et al., 2010, ApJ, 720, 1118
\bibitem[\protect\citeauthoryear{Buchhave et al.}{2012}]{buchhave2012} Buchhave L. A. et al., 2012, Nature, 486, 375
\bibitem[\protect\citeauthoryear{Castelli \& Kurucz}{2004}]{ODF} Castelli F., \&  Kurucz, R. L., 2004, astro-ph/0405087
\bibitem[\protect\citeauthoryear{Collier et al.}{2006}]{collier} Collier Cameron A., Pollacco D., Street R. A. et al. 2006, MNRAS, 373, 799C
\bibitem[\protect\citeauthoryear{Dotter et al.}{2008}]{dotter} Dotter A. et al., 2008, ApJS, 178, 89
\bibitem[\protect\citeauthoryear{Eastman et al.}{2013}]{exofast} Eastman J., Gaudi B.S., \& Agol E., 2013, PASP, 125, 83
\bibitem[\protect\citeauthoryear{Kass et al.}{1995}]{kass} Kass R. \& Raftery A. 1995, Journal of the American Statistical Assosiation, 90, 430, 773-795
\bibitem[\protect\citeauthoryear{Fortney et al.}{2007}]{F2007} Fortney J. J., Marley M. S. \& Barnes J. W. 2007, ApJ, 659, 1661
\bibitem[\protect\citeauthoryear{Fortney et al.}{2008}]{F2008} Fortney J. J. et al., 2008, ApJ, 678, 1419
\bibitem[\protect\citeauthoryear{Jackson et al.}{2005}]{jackson} Jackson B., Greenberg R., \& Barnes R., 2008, ApJ, 678, 1396
\bibitem[\protect\citeauthoryear{Kov\'{a}cs et al.}{2005}]{kovacs1} Kov\'{a}cs G., Bakos G. and Noyes R., 2005, MNRAS, 356, 557
\bibitem[\protect\citeauthoryear{Kov\'{a}cs et al.}{2002}]{kovacs2} Kov\'{a}cs G., Zucker S. \& Mazeh T., 2002, A\&A, 391, 369
\bibitem[\protect\citeauthoryear{Leconte et al.}{2010}]{leconte} Leconte J., Chabrier G., Baraffe I., Levrard B., 2010, A\&A, 516, 64
\bibitem[\protect\citeauthoryear{Lucy \& Sweeney}{1971}]{Lucy} Lucy L. B. \& Sweeney M. A. 1971, ApJ, 76, 544
\bibitem[\protect\citeauthoryear{Maxted et al.}{2015}]{maxted} Maxted P. F. L., Serenelli A. M., \& Sputhworth J., 2015, A\&A, 577, 90
\bibitem[\protect\citeauthoryear{P\'{a}l}{2008}]{pal} P\'{a}l A., 2008, MNRAS, 390, 281
\bibitem[\protect\citeauthoryear{Pollaco et al.}{2006}]{swasp} Pollaco, D. L. et al, 2006, PASP, 118, 1407
\bibitem[\protect\citeauthoryear{Rawlings et al.}{1998}]{Rawlings} Rawlings J. O., Pantula S. G., Dickey D. A. 1998, Applied Regression Analysis: A Research Tool, Second Edition Springer, New York, NY
\bibitem[\protect\citeauthoryear{Southworth et al.}{2009}]{south1} Southworth J. et al., 2009, MNRAS, 396, 1023
\bibitem[\protect\citeauthoryear{Southworth et al.}{2014}]{south2} Southworth J. et al., 2014, MNRAS, 444, 776
\bibitem[\protect\citeauthoryear{Tamuz et al.}{2005}]{tamuz} Tamuz O., Mazeh T. and Zucker S., 2005, MNRAS, 356, 1466
\bibitem[\protect\citeauthoryear{Torres et al.}{2010}]{torres} Torres G., Andersen J. and Gim\'{e}nez A., 2010, A\&ARv, 18, 67
\bibitem[\protect\citeauthoryear{Valenti \& Piskunov}{1996}]{sme} Valenti J. A., \& Piskunov, N., 1996, A\&AS, 118, 595
\bibitem[\protect\citeauthoryear{Vidal-Madjar et al.}{2005}]{vidal} Vidal-Madjar A. et al., 2008, ApJ, 676L, 57

\end{thebibliography}
\end{document}